\documentclass[11pt]{article}
\usepackage{psfig}

\begin{document}

\begin{center}
{\Huge Evolution of the polarization of the optical afterglow of 
the gamma-ray burst GRB 030329}
\end{center}
\bigskip

Jochen Greiner$^*$, Sylvio Klose$^\dag$, Klaus Reinsch$^\ddag$, Hans
Martin Schmid$^\ell$, Re'em Sari$^\$$, Dieter H. Hartmann$^\P$,
Chryssa Kouveliotou$^\S$, Arne Rau$^*$, Eliana Palazzi$^\diamondsuit$,
Christian Straubmeier$^{**}$, Bringfried Stecklum$^\dag$, Sergej
Zharikov$^{\dag\dag}$, Gaghik Tovmassian$^{\dag\dag}$, Otto
B\"arnbantner$^\sharp$, Christoph Ries$^\sharp$, Emmanuel
Jehin$^{\P\P}$, Arne Henden$^{\S\S}$, Anlaug A. Kaas$^{\ddag\ddag}$,
Tommy Grav$^{\$\$}$, Jens Hjorth$^{\ell\ell}$, Holger
Pedersen$^{\ell\ell}$, Ralph A.M.J. Wijers$^\#$, A. Kaufer$^{\P\P}$,
Hye-Sook Park$^{\diamondsuit\diamondsuit}$, Grant
Williams$^{\sharp\sharp}$, Olaf Reimer$^{\#\#}$

\bigskip
\bigskip

\noindent$^*$Max-Planck-Institut f\"ur extraterrestrische Physik, 85741 Garching,
Germany  

\noindent$^\dag$Th\"uringer Landessternwarte, 07778 Tautenburg, Germany

\noindent$^\ddag$Universit\"ats-Sternwarte G\"ottingen, 37083 G\"ottingen, Germany

\noindent$^\ell$Institut f\"ur Astronomie, ETH Z\"urich, 8092 Z\"urich, Switzerland

\noindent$^\$$California Institute of Technology, Theoretical Astrophysics 130-33,
Pasadena, CA 91125, USA

\noindent$^\P$Clemson University, Department of Physics and Astronomy, Clemson, SC
29634, USA

\noindent$^\S$NSSTC, SD-50, 320 Sparkman Drive, Huntsville, AL 35805, USA  

\noindent$^\diamondsuit$Istituto di Astrofisica Spaziale e Fisica
Cosmica, CNR, Sezione di Bologna, 40129 Bologna, Italy

\noindent$^{**}$ I. Physikalisches Institut, Universit\"at K\"oln, 50937 K\"oln,
Germany  

\noindent$^{\dag\dag}$Instituto de Astronomia, UNAM, 22860 Ensenada, Mexico

\noindent$^\sharp$Wendelstein-Observatorium, Universit\"atssternwarte, 81679 M\"unchen,
 Germany

\noindent$^{\P\P}$European Southern Observatory, Alonso de Cordova 3107, Vitacura,
 Casilla 19001, Santiago 19, Chile

\noindent$^{\S\S}$Universities Space Research Association, U.S. Naval
Observatory, P.O. Box 1149, Flagstaff, AZ 86002, USA

\noindent$^{\ddag\ddag}$Nordic Optical Telescope, 38700 Santa Cruz de La Palma, Spain

\noindent$^{\$\$}$University of Oslo, Institute for Theoretical Astrophysics, 0315
Oslo, Norway, and Harvard-Smith\-sonian Center for Astrophysics, Cambridge, MA 02138, USA

\noindent$^{\ell\ell}$Astronomical Observatory, NBIfAFG, University of Copenhagen, 2100
Copenhagen ¯, Denmark

\noindent$^\#$Astronomical Institute Anton Pannekoek, Kruislaan 403, 1098 SJ Amsterdam, The
Netherlands

\noindent$^{\diamondsuit\diamondsuit}$Lawrence Livermore National
Laboratory, University of California, P.O. Box 808, Livermore, CA 94551, USA 

\noindent$^{\sharp\sharp}$MMT Observatory, University of Arizona, Tucson, AZ 85721, USA

\noindent$^{\#\#}$Theoretische Weltraum-und Astrophysik, Ruhr-Universit\"at Bochum, 44780 Bochum, Germany

\bigskip
{\bf The association of a supernova with GRB 030329$^{1,2}$ strongly
supports the collapsar model$^3$ of $\gamma$-ray bursts (GRBs), where
a relativistic jet$^4$ forms after the progenitor star collapses. Such
jets cannot be spatially resolved because of their cosmological
distances. Their existence is conjectured based on breaks in GRB
afterglow light curves and the theoretical desire to reduce the GRB
energy requirements. Temporal evolution of polarization$^{5,6,7}$ may
provide independent evidence for the jet structure of the relativistic
outflow. Small-level polarization ($\sim$1-3\%)$^{8-17}$ has been
reported for a few bursts, but the temporal evolution of polarization
properties could not be established. Here, we report polarimetric
observations of the afterglow of GRB 030329 with high signal-to-noise
and high sampling frequency. We establish the polarization light
curve, detect sustained polarization at the percent level, and find
significant variability. The data imply that the afterglow magnetic
field has small coherence length and is mostly random, probably
generated by turbulence, in contrast with the high polarization
detected in the prompt $\gamma$-rays from GRB 021206$^{18}$. Our results
suggest a different structure and origin of the magnetic field in the
prompt vs. afterglow emission regions.}

\bigskip

GRB 030329 triggered the High Energy Transient Explorer, HETE-II, on
March 29, 2003 (11:37:14.67 UT)$^{19}$. The discovery of the burst
optical afterglow$^{20,21}$ was quickly followed by a redshift
measure\-ment$^{22}$ for the burster of z=0.1685 ($\sim$800 Mpc), thus
making GRB 030329 the second closest long-duration$^{23}$ GRB ever
studied, after GRB 980425$^3$. The proximity of GRB 030329 resulted in
very bright prompt and afterglow emission, leading to the best-sampled
afterglow to date. Detailed optical spectroscopy revealed an
underlying supernova (SN 2003dh)$^{1,2}$ with an astonishing spectral
similarity to SN 1998bw (at z=0.0085, associated with GRB 980425$^3$),
thus strongly supporting the link of long-duration GRBs with
core-collapse supernovae.

\bigskip

The apparent brightness of GRB 030329 also afforded a unique
opportunity to study the temporal evolution of polarization in the
afterglow phase. Previous single measurements found low-level
(1-3\,\%) polarization$^{8-15}$ in optical afterglows, and the only
reports on variable polarization$^{16,17}$ were based on few
measurements with different instruments and modest signal-to-noise. We
have overcome the previous sampling limitations with 31 polarimetric
observations of the afterglow of GRB 030329 obtained with the same
instrumentation (plus few more with different instruments) over a time
period of 38\,days for a total of an unprecedented $\sim$50 hours of
observations with an 8\,m telescope.  For the first time we establish
a polarization light curve for an optical gamma-ray burst afterglow.

\bigskip

We performed relative photometry, and derived from each pair of
simultaneous measurements at orthogonal angles the Stokes parameters U
and Q. In order to obtain the intrinsic polarization of the GRB
afterglow, we have to correct for Galactic interstellar polarization
(mostly due to dust). We performed imaging polarimetry to derive the
polarization parameters of seven stars in the field of GRB 030329, and
obtained an interstellar (dust) polarization correction of 0.45\% at
position angle 155$\deg$. Subtraction of the mean foreground
polarization was performed in the Q/U plane
(Q$_{fp}$=0.0027$\pm$0.0013, U$_{fp}$=-0.0033$\pm$0.0017).  Figure 1
shows an R band image of the GRB 030329 field with the polarization
"vectors" superimposed.

\bigskip

The temporal evolution of the degree and angle of polarization
together with the R band photometry is shown in Figure 2. This figure
demonstrates the presence of non-zero polarization,
$\Pi\sim$0.3-2.5\,\% throughout a 38-day period, with significant
variability in degree and angle on time scales down to hours. Further,
the spectropolarimetric data of the first three nights as well as the
simultaneous R and K band imaging polarimetry during the second night
show that the relative polarization and the position angle are
wavelength independent (within the measurement errors of about
0.1\,\%) over the entire spectral range. These data imply that
polarization due to dust in the host galaxy of GRB 030329 does not
exceed $\sim$0.3\,\%. Dust destruction in the vicinity of the
gamma-ray burster due to the strong radiation field would result in a
monotonically decreasing polarization degree at constant position
angle, which is not observed during the first days. We therefore
conclude that the bulk of the observed variability is intrinsic to the
afterglow.

\bigskip

Figure 2 shows that while the polarization properties show substantial
variability (for which no simple empirical relationship is apparent),
the R band flux is a sequence of power laws. During each of the power
law decay phases the polarization is of order few percent, different
from phase to phase, and variable within the phase, but not in tandem
with the "bumps and wiggles" in the light curve. We observe a
decreasing polarization degree shortly after the light curve break at
$\sim$0.4\,days (as determined from optical$^{21,24}$ and X-ray
data$^{25,26}$). Rapid variations of polarization occur
$\sim$1.5\,days after the burst, and could be related to the end of
the transition period towards a new power law phase starting at
$\sim$1.7\,days. Polarization eventually rises to a level of
$\sim$2\,\%, which remains roughly constant for another two weeks. At
late time the underlying Type Ic SN 2003dh$^{1,2}$ increasingly
contributes to the total light and probably also to the observed
polarization properties. Asymmetries in this type of supernova can
produce polarization of order 1\,\%$^{27,28}$, as we observe towards
the end of our campaign.

\bigskip

GRB  030329 belongs to  a growing  group of  bursts for  which densely
monitored  afterglow   light  curves  show   significant  "bumps"  and
"wiggles"  relative to  a simple  power  law decay  (most notably  GRB
021004  and  GRB  011211).  These  bumps and  wiggles  complicate  the
interpretation  of  the   polarization  properties.  While  the  rapid
decrease in  polarization degree during the first  night is consistent
with the  model predictions$^{5,6,7}$, the position  angle changes are
not,  and thus  it remains  to be  proven whether  or not  these early
polarization data  support the break at  0.4\,days as due to  a jet. A
connection to theoretical models  throughout the entire period covered
in Fig.~2 is  a daunting task. A detailed comparison  of our data with
characteristic features  of various  theoretical models is  beyond the
scope of this paper, and will be presented elsewhere.

\bigskip

In summary, our data constitute the most complete and dense sampling
of the polarization behaviour of a GRB afterglow to date. For GRB
030329 we conclude that the afterglow polarization probably did not
rise above $\sim$2.5\,\% at any time, and that the polarization did
not correlate with the flux. The low level of polarization implies
that the components of the magnetic field parallel and perpendicular
to the shock do not differ by more than $\sim$10\,\%, and suggests an
entangled magnetic field, probably amplified by turbulence behind
shocks, rather than a pre-existing field. This is in contrast with the
high level of polarization detected in the prompt $\gamma$-rays from
GRB 021206$^{18}$ and suggests a different structure and origin of the
magnetic field in the prompt vs. afterglow emission regions. Evolving
polarization properties provide a unique diagnostic tool for GRB
studies, and the extremely complex light curve of the optical
afterglow of GRB 030329 emphasises that measurements should be carried
out with high sampling frequency.

\bigskip

\noindent1. Stanek, K.Z., et al. Spectroscopic discovery of the supernova 2003dh associated with GRB
030329. Astrophys. J. 591, L17-L20 (2003)

\bigskip\noindent2. Hjorth, J., et al. A
very energetic supernova associated with the $\gamma$-ray burst of 29 March
2003. Nature 423, 847-850 (2003)

\bigskip\noindent3. Woosley, S.E., Eastman, R.G.,
Schmidt, B.P. Gamma-ray bursts and type Ic supernova SN
1998bw. Astrophys. J. 516, 788-796 (1999)

\bigskip\noindent4. MŽsz‡ros,
P. Theories of gamma-ray bursts. Ann. Rev. Astron. Astrophys. 40,
137-169 (2002)

\bigskip\noindent5. Sari, R. Linear polarization and proper motion in
the afterglow of beamed gamma-ray bursts. Astrophys. J. 524, L43-L46
(1999)

\bigskip\noindent6. Gruzinov, A. Strongly polarized optical afterglows of
gamma-ray bursts. Astrophys. J. 525, L29-L31 (1999)

\bigskip\noindent7. Ghisellini,
G., Lazzati, D. Polarization light curves and position angle variation
of beamed gamma-ray bursts. Mon. Not. R. Astron. Soc. 309, L7-L11
(1999)

\bigskip\noindent8. Hjorth, J., et al. Polarimetric constraints on the optical
afterglow emission from GRB 990123. Science 283, 2073-2075
(1999)

\bigskip\noindent9. Wijers, R.A.M.J., et al. Detection of polarization in the
afterglow of GRB 990510 with the ESO Very Large
Telescope. Astrophys. J. 523, L33-L36 (1999)

\bigskip\noindent10. Covino, S., et
al. GRB 990510: linearly polarized radiation from a
fireball. Astron. Astrophys.  348, L1-L4 (1999)

\bigskip\noindent11. Rol, E., et
al. GRB 990712: First indication of polarization variability in a
gamma-ray burst afterglow. Astrophys. J. 544, 707-711
(2000)

\bigskip\noindent12. Bj\"ornsson, G., Hjorth, J., Pedersen, K., Fynbo,
J.U. The afterglow of GRB 010222: A case of continuous energy
injection. Astrophys. J. 579, L59-L62 (2002)

\bigskip\noindent13. Masetti, N., et
al. Optical and near-infrared observations of the GRB 020405
afterglow. Astron. Astrophys. 404, 465-481 (2003)

\bigskip\noindent14. Covino, S., et
al. Polarization evolution of the GRB 020405
afterglow. Astron. Astrophys. 400, L9-L12 (2003)

\bigskip\noindent15. Bersier, D., et
al. The strongly polarized afterglow of GRB 020405. Astrophys. J. 583,
L63-L66 (2003)

\bigskip\noindent16. Barth, A., et al. Optical spectropolarimetry of
the GRB 020813. Astrophys. J. 584, L47-L51 (2003)

\bigskip\noindent17. Rol, E., et
al. Variable polarization in the optical afterglow of GRB
021004. Astron. Astrophys. 405, L23-L27 (2003)

\bigskip\noindent18. Coburn, W., and
Boggs, S.E. Polarization of the prompt (-ray emission from the (-ray
burst of 6 December 2002. Nature 423, 415-417 (2003)

\bigskip\noindent19. Vanderspek,
R., Crew, G., Doty, J., Villasenor, J., Monelly, G. GRB 030329
(=H2652): A long, extremely bright GRB localized by HETE WXM and
SCX. GCN Circ. 1997 (2003)

\bigskip\noindent20. Price, P.A., et al. The bright optical
afterglow of the nearby (-ray burst of 29 March 2003. Nature 423,
844-847 (2003)

\bigskip\noindent21. Uemura, M., et al. Structure in the early
afterglow light curve of the (-ray burst of 29 March 2003. Nature 423,
843-844 (2003)

\bigskip\noindent22. Greiner, J., et al. Redshift of GRB 030329. GCN
Circ. 2020 (2003)

\bigskip\noindent23. Kouveliotou, C., et al. Identification of two
classes of gamma-ray bursts. Astrophys. J. 413, L101-L104
(1993)

\bigskip\noindent24. Burenin, R., et al. First hours of the GRB 030329 optical
afterglow. Astron. Lett.  29, 9, 1-6 (2003)

\bigskip\noindent25. Marshall, F.E.,
Markwardt, C., Swank, J.H. GRB 030329: fading X-ray afterglow with
RXTE.  GCN Circ. 2052 (2003) 

\bigskip\noindent26. Tiengo, A., et al. The X-ray
afterglow of GRB 030329. Astron. Astrophys., in press;
astro-ph/0305564 (2003)

\bigskip\noindent27. Wang, L., Howell, D. A., Hšflich, P.,
and Wheeler, J. C. Bipolar supernova explosions. Astrophys. J. 550,
1030-1035 (2001)

\bigskip\noindent28. Leonard, D. C., Filippenko, A. V., Chornock, R.,
and Foley, R. J. Photospheric-phase spectropolarimetry and
nebular-phase spectroscopy of the peculiar type Ic supernova
2002ap. Publ. Astron. Soc. Pac. 114, 1333-1348
(2002)

\bigskip
\bigskip

\noindent Acknowledgements
 
This work is primarily based on observations collected at ESO, Chile,
with additional data obtained at the German-Spanish Astronomical
Centre Calar Alto, operated by the Max-Planck-Institute for Astronomy,
Heidelberg, jointly with the Spanish National Commission for
Astronomy, the NOT on La Palma, Canary Islands, and the Observatorio
Astronomico National, San Pedro, Mexico. We are grateful to the staff
at the Paranal, Calar Alto and NOT observatories, in particular
A. Aguirre, M. Alises, S. Hubrig, A.O. Jaunsen, C. Ledoux, S. Pedraz,
T. Szeifert, L. Vanzi and P. Vreeswijk for obtaining the service mode
data reported here.



\clearpage
\begin{figure}
\vspace{-1.0cm}
\begin{center}
\psfig{figure=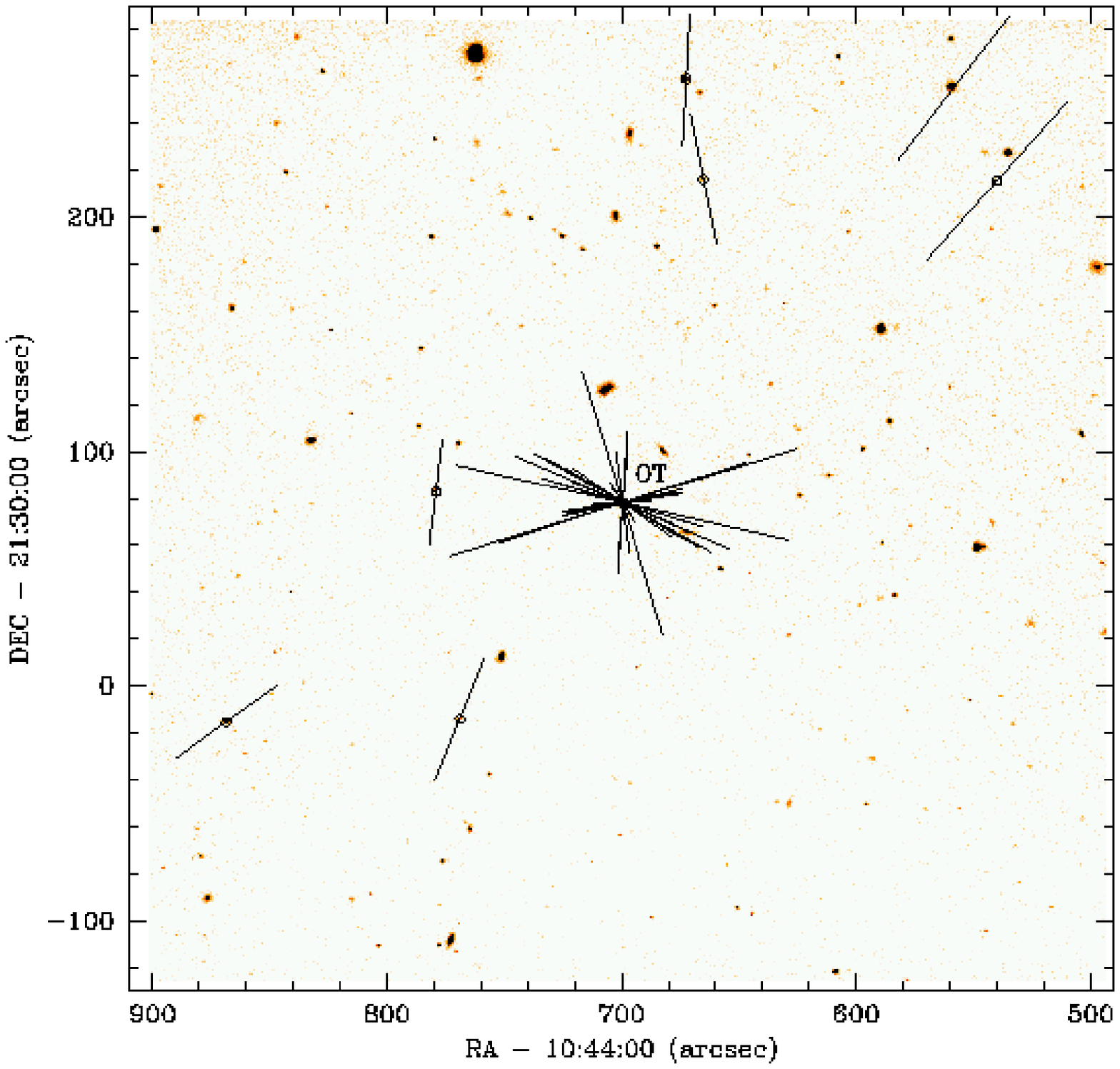,width=14.5cm}%
\end{center}
\end{figure}

\vspace{-1.0cm}
{\bf Figure 1:} R band image of the field centred on
the optical afterglow of GRB 030329. The 27 FORS1 (Focal Reducer and
Low Dispersion Spectrograph at VLT/Antu) measurements are shown as
"vectors" where the length is a measure of the polarization degree,
and the orientation indicates the position angle. While the afterglow
varies in degree as well as angle, the polarization of seven field
stars is constant within 0.1\% in polarization degree and 1.5 degree in
position angle throughout the 38 days. Linear polarization was
measured from sets of exposures with different retarder plate position
angles. Imaging polarimetry was obtained during the first four nights
(when the afterglow was brighter than 17th mag) from sixteen different
retarder plate angles, and from eight angles thereafter. Since the
FORS1 polarization optics allows determination of the degree of
polarization to an accuracy of <3x10-4 and of the polarization angle
to $\sim$ 0.2 deg, we consider the above variance of the field stars to
represent the systematic error over the 38-day time
period. Observations of polarimetric standard stars reproduced their
tabulated values within 5\%. The FORS1 retarder plate zero point angle
of -1.2 degrees was subtracted from the polarization angle. The
position angle has a systematic uncertainty of ±1.5 deg, which was
added in quadrature to the statistical errors.

\clearpage
\begin{figure}
\vspace{-1.0cm}
\begin{center}
\psfig{figure=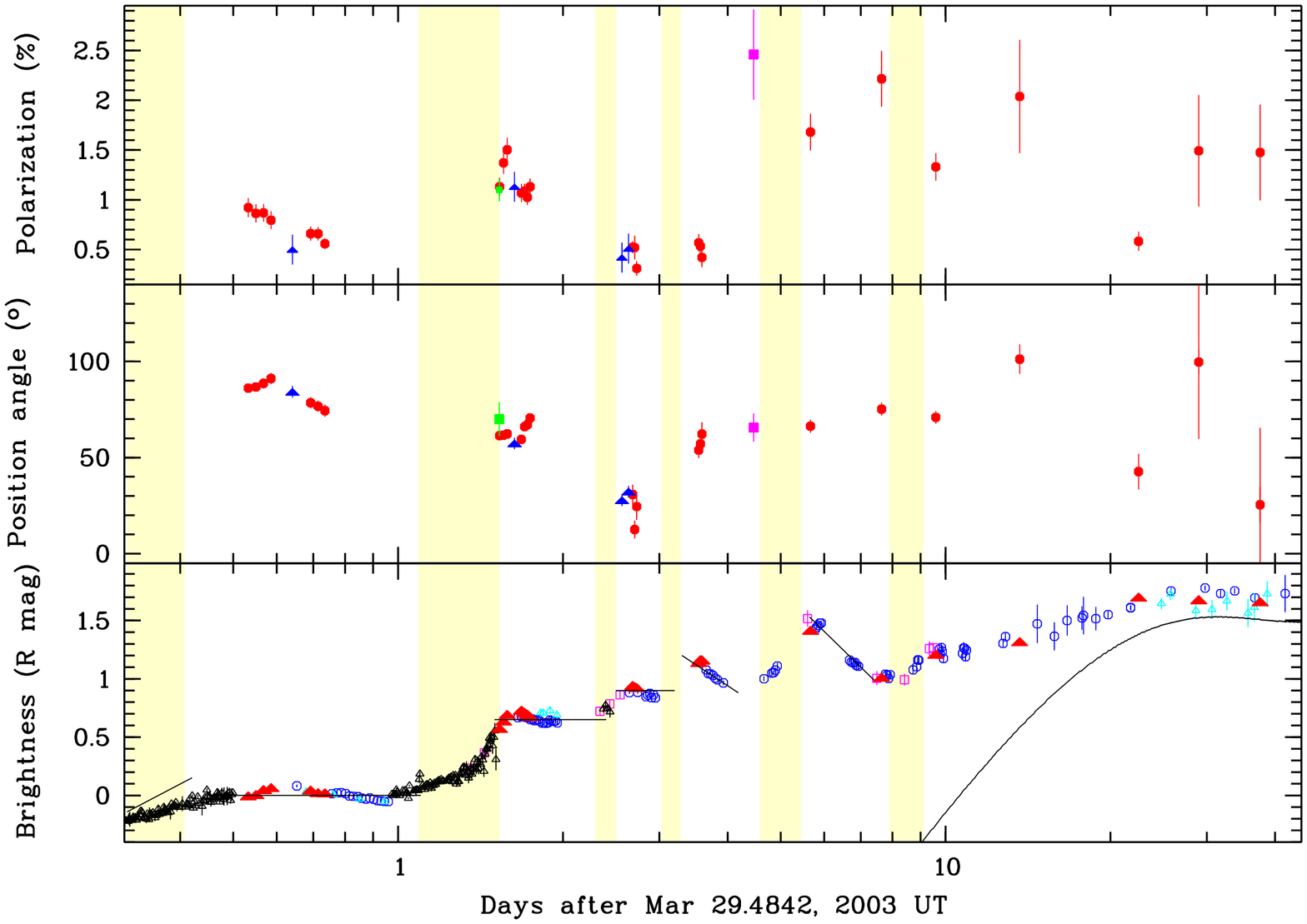,width=15.5cm,angle=-90}%
\end{center}
\end{figure}

\vspace{-1.0cm}
{\bf Figure 2:} Evolution of the polarization during the first 38 days. Top
and middle panels show the polarization degree in percent and the
position angle in degrees. The red data points are from imaging
polarimetry with FORS1/VLT. Spectropolarimetry (blue symbols, 700-800
nm range) was performed during the first three nights. The green
points were obtained with CAFOS at the 2.2m telescope at Calar Alto on
March 30/31. The magenta data point was obtained with AFOSC at the
2.56m NOT telescope on 2003 April 2. The bottom panel shows the
residual R band light curve after subtraction of the contribution of a
power law t-1.64 describing the undisturbed decay during the time
interval 0.5-1.2 days after the GRB (i.e., after the early break at
0.4 days), thus leading to a horizontal curve. The symbols correspond
to data obtained through: the literature (black), the 1m USNO
telescope at Flagstaff (blue), the OAN Mexico (cyan), and FORS1/VLT
(red). Lines indicate phases of power law decay, with the first one
from early data21 (not shown). Yellow bars mark re-brightening
transitions. Contributions from an underlying supernova (solid curved
line) do not become significant until $\sim$10 days after the GRB.

\end{document}